\documentclass[sigconf]{acmart}

\AtBeginDocument{%
  \providecommand\BibTeX{{%
    \normalfont B\kern-0.5em{\scshape i\kern-0.25em b}\kern-0.8em\TeX}}}

\setcopyright{acmlicensed}
\copyrightyear{2020}
\acmYear{2020}
\acmDOI{10.1145/3379597.3387499}

\acmConference[MSR '20]{17th International Conference on Mining Software
Repositories}{October 5--6, 2020}{Seoul, Republic of Korea}
\acmBooktitle{17th International Conference on Mining Software Repositories
(MSR '20), October 5--6, 2020, Seoul, Republic of Korea}
\acmPrice{15.00}
\acmISBN{978-1-4503-7517-7/20/05}

\begin{document}

\title{A Complete Set of Related Git Repositories Identified via Community Detection Approaches Based on Shared Commits}

\author{Audris Mockus}
\affiliation{%
  \institution{The University of Tennessee}
  \city{Knoxville}
  \state{Tennessee}
}
\email{audris@mockus.org}

\author{Diomidis Spinellis}
\orcid{0000-0003-4231-1897}
\affiliation{
    \institution{Athens University of Economics and Business}
    \city{Athens}
    \country{Greece} 
}
\email{dds@aueb.gr	}

\author{Zoe Kotti} 
\affiliation{
    \institution{Athens University of Economics and Business}
    \city{Athens}
    \country{Greece}
}
\email{zoekotti@hotmail.com}
 
\author{Gabriel John Dusing}
\affiliation{%
  \institution{The University of Tennessee}
  \city{Knoxville}
  \state{Tennessee}
}
\email{gdusing@vols.utk.edu}


\begin{abstract}
  In order to understand the state and evolution of the entirety of
  open source software we need to get a handle on the set of
  distinct software projects. Most of open source projects presently
  utilize Git, which is a distributed version control system
  allowing easy creation of clones and resulting in numerous
  repositories that are almost entirely based on some parent
  repository from which they were cloned. Git commits are unlikely
  to get produce and represent a way to group cloned
  repositories. We use World of Code infrastructure containing
  approximately 2B commits and 100M repositories to create and share
  such a map. We discover that the largest group contains almost 14M
  repositories most of which are unrelated to each other. As it
  turns out, the developers can push git object to an arbitrary
  repository or pull objects from unrelated repositories, thus
  linking unrelated repositories. To address this, we apply Louvain
  community detection algorithm to this very large graph consisting
  of links between commits and projects. The approach successfully
  reduces the size of the megacluster with the largest group of
  highly interconnected projects containing under 400K
  repositories. We expect that the resulting map of related projects
  as well as tools and methods to handle the very large graph will
  serve as a reference set for mining software projects and other
  applications. Further work is needed to determine different types
  of relationships among projects induced by shared commits and
  other relationships, for example, by shared source code or similar
  filenames.
\end{abstract}
\vspace{-.2in}

\maketitle
\section{Introduction}
\vspace{-.03in}
While study of individual software projects has been ongoing for
some time, relatively less effort has been spent on studying groups
of projects even though numerous benefits of understanding groups of
projects exist. Extensive motivation for investigating groups of
projects was stated in, for
example~\cite{HMPQ10,hackbarth2016assessing}. Furthermore, several
attempts at creating an infrastructure for such studies have been
reported~\cite{di2017software,czerwonka2013codemine,Dyer-Nguyen-Rajan-Nguyen-13,gousios2012ghtorrent}. Here
we focus on an aspect of apparently simple, but very hard to address
challenge of identifying related repositories. By related repositories we mean 
repositories that are ``developed for the same project/component.'' Unrelated 
repositories are, thus, are not intending to merge their code to a
single project. For example, while most GitHub forks are not meant to be
independent projects, but some are. Once we go beyond GitHub, often
no information about relatedness is available. For, example,
searching for relevant code or looking for a project to join, the
massive number of the repositories would waste time and cause confusion. 

Simply stated, if we obtain data from two distinct git repositories,
should we treat it as belonging to a single project or as belonging
to two unrelated projects. This is an important question since many
projects have tens of thousands of forks that often contain no or
very little original content as the fork was created just to submit
a pull request and may not have even been used for that purpose.
Some relief can be found for projects on GitHub, where forked
projects can be identified via GitHub API. No such information is
available for other projects and projects that were not forked via
GitHub API will be missed. Having a reference list of related
projects can provide a massive help to Mining Software Repositories
community by providing a common basis that everyone can use to
count, identify, sample from, or analyze projects. Our operational
definition of a set of repositories representing an independent
project is that all these repositories share the objective to work
on the same project. For example, a fork created to submit a pull
request or to fix a bug that, for all practical purposes, is
expected to be eventually fixed upstream can be illustrated by the
repositories Debian distribution uses to keep track of upstream
packages.  They are used to ensure that everything compiles and can
be installed together for the distribution, but are not intended to
maintain the upstream project.  This also includes cases where a
project may maintain another project within its own repository, but
not with the purpose of developing it, just to avoid potential
incompatibilities that may occur due to differences in development
schedule. Hard forks, on the other hand, would indicate a desire to
develop the project independently and should be considered as
separate projects.

While there are many ways to identify related projects, here we
focus on a single approach: linking projects sharing at least one
commit.  Git commits are based on Merkle Tree and no two commits are
likely to be produced independently. For example, the
initial commit to a repo creating an empty README.md file and done
at exactly the same time (up to a second) in the same timezone, by
two developers having identical credentials would result in an identical commit. 
However, as a distributed VCS, git makes it easy to create clones (via git clone
or through GitHub fork button) and resulting in numerous
repositories that are intended to be distributed copies of the code
used in the same project. This feature of Git that enables
distributed collaboration also results in numerous clones of the
original repository. Furthermore, GitHub introduced single-click way
to fork (in essence to clone) a repository on GitHub and use it to
create patches (pull requests) for the origin of that fork. This
further increased the number repositories related to popular
projects.

As noted above, it is virtually impossible to produce independently
identical commits in normal development (see a potential example
above), so it would appear that projects sharing a commit are
related. Here we do not consider other types of related projects
where the version history was not shared and only the source code
has been copied. Such projects can not be identified via shared
commits and are a subject of further work, for example by comparing
the blobs shared among the projects or the directory structure of
the source code~\cite{CM,HM08,zzm14}.

To apply the approach we utilize the infrastructure provided by the
World of Code~\cite{woc19}. Specifically, we use version Q of the
data and obtain commit to project relationships. As described in WoC
tutorial~\cite{woctut}, the data is stored in 32 databases containing
a full list of pairs between commits and projects in which these
commits were found (c2pFullQXX.s). This commit to project graph has
a total of 99,154,451,345 links between 116265607 projects and
1868632121 commits. We handle the scale of the problem as described
in Section~\ref{s:link}, i.e., by solving a sequence of smaller
problems and using the results to solve the larger problem.
The largest group of repositories (we use words ``project'' and
``repository'' interchangeably here) has almost 14M projects and not
all of them appear to be closely related.

We, therefore identify some of the reasons for such outcome
(projects that fetch from or push to repositories of unrelated
projects) and propose and implement alternative operationalizations
of related projects. These involve the attempt to identify and
remove problematic projects or commits, simplification of the
problem by reducing the number of projects by using explicit fork
identification in GitHub, and using Louvain community detection
algorithm to separate connected but unrelated projects.

Our results provide an operationalizaton of related projects for
open source projects utilizing git version control system obtained
from WoC infrastructure. In addition to projects related by a shared
commit, we also provide the ultimate parent from forking
relationships for GitHub forks and the groups defined by Louvain
community detection algorithm.

The remainder of the paper describes data sources in
Section~\ref{s:sources}, the approach to link the projects in
Section~\ref{s:link}, the approach to eliminate problematic projects
and commits in Section~\ref{s:bad}, the community detection approach
described in Section~\ref{s:comm}, and summary in
Section~\ref{s:summ}.
\vspace{-.03in}
\section{Data Source: World of Code}\label{s:sources}
\vspace{-.03in}
The World of Code~\cite{woc19} infrastructure prototype was created
to support developing theoretical, computational, and statistical
frameworks that discover, collect, and process FLOSS operational
data and construct FLOSS supply chains (SC), identify and quantify
its risks, and discover and construct effective risk mitigation
practices and tools. That prototype stores the huge and rapidly
growing amount of data in the entire FLOSS ecosystem and provide
basic capabilities to efficiently extract and analyze that data at
that scale. WoC's primary focus is on types of analyses that require
global reach across FLOSS projects.

In a nutshell, WoC is a software analysis pipeline starting from
discovery and retrieval of data, storage and updates, and
transformations and data augmentation necessary for analytic tasks
downstream. In addition to storing objects from all git repositories
WoC also provides relationships among them.  For the purpose of this
analysis we only use a single relationship from WoC: commit to
project map that lists all commit project pairs.  WoC has two
interfaces: one optimized for random access and another for
processing the entirety of the collection. We chose the second due
to need to obtain the entirety of the commit to project links. WoC
splits each relationship into 32 databases. Specifically, the c2p
(commit to Project) database is split based on 5 bits of the first
byte of commits Sha1. Thus we naturally have 32 smaller datasets to
analyze. Randomness of Sha1 ensures that each of these databases
represents the entire collection.

WoC data is versioned with the latest version labeled as Q and
containing 7204111388 blobs, 1868632121 commits, 7596825726 trees,
16172556 tags, 116265607 projects (distinct repositories), and
38142898 distinct author IDs. WoC has collected that data during
November and December of 2019. For more information please consult
WoC website~\cite{wocurl}. The proposed grouping into the related
projects  
produced 66532614 such clusters with the largest cluster containing  
354920 repositories (miranagha/js).

We also use fork parent data obtained from
GHTorrent~\cite{gousios2012ghtorrent} and, for GitHub projects not
present in GHTorrent, we retrieved using GitHub GraphQL
API~\cite{ghapi}.  Please note that that GitHub forks may have their
own forks. For each project we obtain the ultimate parent: that is if
the parent has a parent, we continue until the repository is no
longer fork.
\vspace{-.03in}
\section{Linking projects by shared commits}\label{s:link}
\vspace{-.03in}
As noted above, we distribute the computational load over the 32
databases listing commit/project pairs. Since data in these lists
are sorted we simply need to group projects that share the same
commit. Commits belonging to a single project can be ignored as they
will not provide a link among projects. The result of the first pass
over each of the 32 databases is a list of lines each listing two
or more projects linked by a commit (for more detail please see ssc-oscar/forks/README.md). We then encode each line
representing a group of N projects as $N-1$ links linking the first
project to the remaining ones. The resulting graph is used to
produce cliques (connected components) via C$++$ Boost library~\cite{boost}\footnote{Specifically connected\_components function from
``boost/graph/connected\_components.hpp''}.  The resulting
components from each of the 32 databases are then combined into a
single graph and the same library is used to produce the overall
components. The largest components are shown in Table~\ref{tab:mega}.
\begin{table}[htb]
    \centering
    \begin{tabular}{c|c}
    Member Count & Name \\
 13,912,612 & grr \\
  28,193 & rh24/parrot-ruby \\
  17,267 & kvignali/arel-lab \\
  16,181 & hmagph/ui$-$ \\
  16,170 & 54/996\.ICU \\
  10,541 & mil/kb \\
  10,218 & bloomni/aa \\
   9,911 & f0/rkt \\
    \end{tabular}
    \caption{The largest groups of related repositories}
    \label{tab:mega}
    \vspace{-.4in}
\end{table}
Names of the clusters are chosen by selecting a repository from the
cluster that has the shortest name that and is first in the
alphanumeric order.  This cluster name is provided as the second
column of the provided map, where the first column lists all
116,265,607 projects and the linking produces 61,921,909 distinct
clusters unconnected by commits.
The first mega-cluster exceeds the next one by almost three
magnitudes and is clearly undesirable as it packs more than 10\% of
all projects and groups together what appear to be rather unrelated
projects.
\vspace{-.03in}
\section{Removing bad projects and commits}\label{s:bad}
\vspace{-.03in}
Given less than ideal outcome obtained in Section~\ref{s:link}, we
have spent some time investigating the reasons behind that
outcome. Specifically we identified at least two kind of
repositories that give rise to such a mega-cluster. First, it
appears that some projects are used in what appears to be simply a
backup storage. Since any developer who has a permission to write to
a repository can push git objects to it from any unrelated
repository, this feature may have been used by some developers to
use cloud git version control systems simply to back up their
work. Examples of such repositories include
``docker-library/commit-warehouse'' and
``devillnside/AcerRecovery.'' The second class of problematic
repositories appears to include repositories that contain version
history from multiple independent projects that are used to build a
single project, for example, ``bloomberg/chromium.bb'' that contains
commit history from independent projects such as libdrm and
FFmpeg. A simple attempt to remove such projects manually did not
give great results as after one of such problematic projects was
removed, there we hundreds of others that remained leaving the size
of the mega-cluster stubbornly high. After eliminating a large set
of potentially problematic projects (listed in the code as an
associative array bad Projects) and also removing potentially
problematic commits (commits that span more than one thousand
projects), we still had a formidable mega-cluster containing
9,626,594 projects and 65,591,526 groups of unrelated projects.
\vspace{-.03in}
\section{Community Detection}\label{s:comm}
\vspace{-.03in}
Research on large graphs has produced a number of algorithms that
detect communities: groups that interact (have more links) among
themselves than across groups. Such algorithms tend to be much more
time consuming than the arguably simplest connected subset detection
algorithm we used in Section~\ref{s:link}. More importantly, it is
not clear how to combine the results from multiple runs of the
algorithm on different subsets of commits as we did in that
section. We, therefore, tried to simplify the problem in two
ways. First, we reduce the number of distinct projects by using
information obtained from GitHub fork API and substituting project
name by 
its ultimate parent.  Second, we reduce the number of commits by
considering all commits touching the same subset of projects as a
single commit. That resulted in 141,53,282 groups (hyperlinks) of
projects representing a minimum of 13,82,233,820 links involving two
projects.

After this preparation the
resulting graph was then analyzed using iGraph package in
R~\cite{R}. It was necessary to read and add links to the graph in
chunks in order to avoid creating a long vector (iGraph can not
presently handle R's long vector). Louvain's algorithm implementation
in R was used, specifically, the function
``cluster\_louvain''~\cite{louvain}.
Louvain's algorithm optimizes modularity. It was chosen for its speed--
spectral clustering requires $O(n^3)$ operations while Louvain is nearly linear on the sparse graph we have.

The resulting set of groups (we use groups, components, and clusters
interchangeably) appears to be much more reasonable with
the three largest groups representing what appear to be legitimately related groups of
projects involved in language tutorials (miranagha/js), github.io templates of creating a static github website/personal CV
(6101/-) programming assignments (ykgm/R), datasharing templates
(jkwonl/test), linux kernel for mobile mods (aosp/oz), bootstrap
(UCF/50), configuration files (rdp/a), and spring framework (maiyy/-).
\vspace{-.03in}
\begin{table}[htb]
    \centering
    \begin{tabular}{c|c}
    Member Count & Name \\
354920&miranagha/js\\
333645&6101/-\\
241893&ykgm/R\\
211538&jkwonl/test\\
179315&aosp/oz\\
101988&UCF/50\\
94160&rdp/a\\
89602&maiyy/-\\
    \end{tabular}
    \caption{The largest groups of related repositories using Louvain community detection}
    \label{tab:largest}
    \vspace{-.4in}
\end{table}
\section{Data Overview}
\vspace{-.03in}
We provide ``ultimateMap2.s'' where the first column is the repo
name transformed with the first `/' replaced by `\_', and the
`github.com/' removed for GitHub repositories.  It was produced by
assigning the result from running community detection mentioned
previously to assign a cluster name which represents an
independently developed project to all repositories in World of
Code.

We also provide `ghForks.gz' which is produced by identifying, for each forked repository, its ultimate parent. If a parent is a fork itself, 
find its parent, and so on, until it is not a fork.
\vspace{-.03in}
\section{Evaluation}
\vspace{-.03in}
To measure the accuracy of the community detection algorithm we rely
on the incomplete list of ultimate parent repositories at the time
we performed the clustering calculation. Over the weeks during which
we were doing the clustering, we also retrieve information
using GitHub API on whether or not the project was a fork and, if
so, what was the parent. Over 15M projects from WoC could not be
found in the ghTorrent extract we used. Due to throtling of GitHub
API the process of obtaining fork parents is very slow. Over the
period we did the computation, we were able to retrieve fork parent
information for only approximately one million GitHub repositories.
By the time of writing we have collected fork information on
1,652,872 repositories that was not available for the community
detection analysis described in this section. We used that
information to determine if the community detection approach was
able to group these repositories to the corresponding ultimate
parents in this new extract. Of these, only 32,082 or 1.9\% were not
placed in the more than one group (represented by the ultimate
parent). Of these incorrectly split, most
(9,245) were in the octocat/Spoon-Knife, which is a test repository for
developers to practice using git. Also, only a tiny percentage of
repositories were separated from the main group, for example only
one repository was split from spring-projects/spring-boot (see
Table~\ref{tab:errors}.
\begin{table}[htb]
    \centering
    \begin{tabular}{c|c|c}
    In split & lrgst grp & Parent fork \\
    9245 & 9222 & octocat/Spoon-Knife \\
    2717 & 2684 & rdpeng/ProgrammingAssignment2 \\
    1957 & 1936 & rdpeng/ExData\_Plotting1 \\
    1046 & 1045 & spring-projects/spring-boot 
    \end{tabular}
    \caption{Split fork parents with most repositories}
    \label{tab:errors}
    \vspace{-.3in}
  \end{table}   
As shown in the table the most repositories in forks that were 
split, occurred in training repositories and with only a few 
repositories not in the primary group.

We also compare our approach to the competing approach
described~\cite{diomidis}.  Specifically, the competing approach
provides groups for 10,649,348 repositories and the set of
repositories that we could match was 8157317. Assuming the competing
approach as the gold standard, our approach splits 100,300 of the
2,036,117 groups (5\%) in the competing approach.  Conversely,
assuming our approach as the gold standard, the competing approach
splits 44,357 out of 2,124,711 (2\%) of the groups we detect using
our algorithm. 

Inspecting the largest discrepancies, our approach produces 629
groups for the torvalds/linux group of the competing approach and
competing approach splits aosp/oz group (kernel mobile mods)
produced by our algorithm into 1,245 groups.  All data and code needed
to reproduce the results are in github.com/ssc-oscar/forks.
\vspace{-5.5pt}
\section{Limitations}\label{s:lim}
\vspace{-5.5pt}
It is important to note that we are not trying to solve the problem
of identifying all related project, just ones that are related via
code commits. We are also not investigating finer types of
relationships, for example, light forks done for a single pull
request vs hard forks where projects evolve independently or all
shades of grey in between. Other approaches may be more suitable (or
used in combination with shared commit methods) for that. For
example, shared code, amount of independent evolution, etc.

We utilize WoC data collection with all associated limitations of
using that repository and described there~\cite{woc19}.

The accuracy of our approach is not easy to establish. While we rely
on explicitly specified GitHub forks in the community detection
step, these may involve cases where the forked projects are
developed independently with no intention to merge. We were able to
reproduce the explicitly specified forks with high accuracy,
however.

Some of the projects in WoC may have been renamed and the new
project may have the name of the old project but an entirely
different content.  Identifying and eliminating such projects would
help improve the accuracy of the community detection algorithms.

While our approach appears to provide sensible groups of projects,
it may be further improved by experimenting with different community
detection algorithms and by weighting links in a different manner.

While there is no particular reason to expect that
modularity-focused algorithms should work on a problem involving the
discrimination of forks and non-forks, it is not unreasonable to
assume that unusual patterns of using git such that objects from
unrelated projects are shared within a single repository, would
appear as anomalies and thus be eliminated as spurious links between
groups of legitimate forks.
\vspace{-5.5pt}
\section{Summary}\label{s:summ}
\vspace{-5.5pt}
The main purpose of this work is to demonstrate the feasibility of
solving the problem of finding groups of repositories representing independent projects 
on a global scale with a high
scale of automation, and to share the resulting dataset with the
research community for further improvement. We also hope that the
resulting map will be incorporated into WoC and other
infrastructures such as BoA~\cite{Dyer-Nguyen-Rajan-Nguyen-13} and
SoftwareHeritage~\cite{di2017software} to further simplify sampling,
counting, and statistical analysis of the open source projects.

Specifically, we discover that a direct application of
commit-sharing resulted in the largest group containing almost 14M
repositories.  This happened because the developers can push git
objects to an arbitrary repository and pull objects from unrelated
repositories into their repository, thus linking unrelated
repositories. We attempted to eliminate such problematic repositories
with limited success until we applied Louvain community detection
algorithm.  The approach successfully reduces the size of the
mega-cluster with the two largest groups of highly interconnected
projects containing approximately 100K repositories that all appear
to be closely related.
As future work, it might be worth considering ways to apply
time-series methods to observe graph behavior over time.
In~\cite{Sikdar2016}, a prediction framework is presented for
certain graph parameters, e.g. modularity or average
degree. Reducing the size of the graph in question may also be
helpful which~\cite{abukhzam2018} suggests how this might be done.
Underpinning this approach is that not all edges of the graph are
necessary to draw conclusions, and by embedding the graph in a
metric space, certain edges close together in some sense can be
treated as a single edge.

We expect the tools that the resulting map of related projects as
well as tools and methods to handle the very large graph will serve
as a reference set for mining software projects and other
applications. Further work, however, will be required to determine
the different types of relationships among projects induced by
shared commits and other relationships, for example, by shared
source code or similar filenames.

The work has been partially supported by the following NSF awards:
CNS-1925615, IIS-1633437, and IIS-1901102


\bibliographystyle{plain}
\bibliography{bibliography,audris,all,bib}

\end{document}